\def\ts     {\thinspace}
\def\kms    {\ifmmode{{\rm \ts km\ts s}^{-1}}\else{\ts km\ts s$^{-1}$}\fi}
\def\msol   {\ifmmode{{\rm M}_{\odot}}\else{M$_{\odot}$}\fi}
\def\lsol   {\ifmmode{{\rm L}_{\odot}}\else{L$_{\odot}$}\fi}
\def\zsol   {\ifmmode{{\rm Z}_{\odot}}\else{Z$_{\odot}$}\fi}
\def\etal   {{\rm et\ts al.}}
\def\aco    {\ifmmode{^{12}{\rm CO}(J\!=\!1\! \to \!0)}\else{$^{12}${\rm CO}($J$=1$\to$0)}\fi}
\def\bco    {\ifmmode{^{12}{\rm CO}(J\!=\!2\! \to \!1)}\else{$^{12}${\rm CO}($J$=2$\to$1)}\fi}
\def\cco    {\ifmmode{^{12}{\rm CO}(J\!=\!3\! \to \!2)}\else{$^{12}${\rm CO}($J$=3$\to$2)}\fi}
\def\ci     {\ifmmode{{\rm C}{\rm \small I}}\else{C\ts {\scriptsize I}}\fi}
\def\hi     {\ifmmode{{\rm H}{\rm \small I}}\else{H\ts {\scriptsize I}}\fi}
\def\hh     {\ifmmode{{\rm H}_2}\else{H$_2$}\fi}
\def\cone {\ifmmode{{\rm C}{\rm \small I}(^3\!P_1\!\to^3\!P_0)}
     \else{C\ts {\scriptsize I}{\small$(^3\!P_1\!\to\,^3\!P_0)$}}\fi}
\def\ctwo {\ifmmode{{\rm C}{\rm \small I}(^3\!P_2\!\to\,^3\!P_1)}
     \else{C\ts {\scriptsize I}{\small$(^3\!P_2\!\to\,^3\!P_1)$}}\fi}
\def\cij {\ifmmode{{\rm C}{\rm \small I}\,(^3P_i\to^3P_j)}\else{C\ts {\scriptsize I}\,{\small$(^3P_i\to^3P_j)$}}\fi}
\def\cii    {\ifmmode{{\rm C}{\rm \small II}}\else{C\ts {\scriptsize II}}\fi}
\def\tex {\ifmmode{{T}_{\rm ex}}\else{$T_{\rm ex}$}\fi}
\def\tmb {\ifmmode{{T}_{\rm mb}}\else{$T_{\rm mb}$}\fi}
\def\tkin {\ifmmode{{T}_{\rm kin}}\else{$T_{\rm kin}$}\fi}
\def\microns {\ifmmode{\mu{\rm m}}\else{$\mu$m}\fi}
\def\nhh   {\ifmmode{n({\rm H}_2)}\else{$n$(H$_2$)}\fi}
\documentclass[twocolumn]{aa}
\usepackage{graphicx}
\begin{document}
 \title{Atomic carbon at redshift~$\sim$2.5 
  %\thanks{based on observations with the IRAM 30m telescope}
  }

   \author{A. Wei\ss
          \inst{1}
          \and
          D. Downes
          \inst{2}
          \and
          C. Henkel
          \inst{3}
          \and
          F. Walter
          \inst{4}
          }

   \institute{IRAM, Avenida Divina Pastora 7, 18012 Granada, Spain
         \and
             IRAM, 300 rue de la Piscine, 38406 St-Martin-d'H\'eres, France
         \and
             MPIfR, Auf dem H\"ugel 69, 53121 Bonn, Germany
         \and
             MPIA, K\"onigstuhl 17, 69117 Heidelberg, Germany 
             }

   \date{}

   \abstract{Using the IRAM 30m telescope we detected the lower fine structure line
     of neutral carbon (\cone, $\nu_{\rm rest} = 492$\,GHz) towards 
     three high--redshift sources: IRAS FSC\,10214 ($z=2.3$), SMM\,J14011+0252 ($z=2.5$) and H1413+117
     (Cloverleaf quasar, $z=2.5$). SMM\,J14011+0252 is the first
     high--redshift, non--AGN source in which \ci\ has been
     detected. The \cone\ line from FSC\,10214 is almost an order of magnitude weaker
     than previously claimed, while our detection in the
     Cloverleaf is in good agreement with earlier observations.
     The \cone\ linewidths are similar to the CO widths, indicating
     that both lines trace similar regions of molecular gas on
     galactic scales. Derived \cone\ masses for all three objects are
     of order few\,$\times10^7$\,\msol\ and the implied \cone/\cco\ 
     line luminosity ratio is about 0.2. This number is
     similar to values found in local galaxies. We derive a \ci\ abundance of $5\times10^{-5}$ which implies
     significant metal enrichment of the cold molecular gas at redshifts
     2.5 (age of the universe 2.7\,Gyr). We conclude that the physical properties of systems at large
     lookback times are similar to today's starburst/AGN environments.

   }

   \maketitle
%
%________________________________________________________________

\section{Introduction} 
Investigating the dense cool interstellar medium in high redshift galaxies
is of fundamental importance for our understanding of the early phases of galaxy
formation and evolution. Recent studies have revealed the presence of
huge molecular gas quantities ($>10^{10}$\,\msol) in distant objects, 
which show that the gas reservoirs for powerful starbursts are present in 
the early universe (for a review see Carilli \etal\ \cite{carilli04}). So far, molecular
gas has been detected in about 25 sources at $z>2$, out to a redshift of
$z= 6.4$ (Walter \etal\ \cite{walter03}, Bertoldi \etal\ \cite{bertoldi03}).

\noindent The brightest and therefore most common tracer for molecular gas is
carbon monoxide (CO). Multi--transition studies of CO lines can be
used to constrain the physical properties of the molecular gas.
However, the interpretation of optically thick CO lines is not
straightforward and requires detailed modeling (e.g., radiative
transfer codes). To constrain these models, observations of faint
optically thin lines (such as e.g., $^{13}$CO) are needed
(e.g. Wei\ss\ \etal\ \cite{weiss01}) which are,
however, inaccessible given current limitations in sensitivity. 

\noindent Observations of neutral carbon (\ci) can help to circumvent these
problems.  Because the 3P fine-structure system of carbon forms a
simple three-level system, detections of {\it both} optically thin
carbon lines, \cone\ (492\,GHz) and \ctwo\ (809\,GHz), in principle
allow us to derive the excitation temperature, neutral carbon column
density and mass independently of any other information (e.g. Stutzki
\etal\, \cite{stutzki97}, Wei\ss\ \etal\ \cite{weiss03}).

\noindent Although the atmospheric transmission is poor at the \ci\ rest frequencies, a
number of studies of neutral carbon have been carried out in 
molecular clouds of the galactic disk, the galactic center, M82 and
other nearby galaxies (e.g., White \etal\ \cite{white94}; Stutzki
\etal\ \cite{stutzki97}; Gerin \& Phillips \cite{gerin00};
Ojha \etal\ \cite{ojha01}; Israel \& Baas \cite{israel02}; Schneider
\etal\ \cite{schneider03}, Kramer \etal\ \cite{kramer04}). These
studies have shown that \ci\, is closely associated with
the CO emission independent of the environment.  Since the critical
density for the \cone\ and \aco\ lines are both $n_{\rm cr} \approx
10^3\,{\rm cm}^{-3}$ this finding suggests that the transitions arise
from the same volume and share similar excitation temperatures
(e.g. Ikeda \etal\, \cite{ikeda02}).

\noindent In this letter we report on the detection of the lower fine structure
line of atomic carbon, \cone, towards the three strongest molecular
line emitters at redshifts $>2$ currently known: IRAS FSC\,10214 
($z=2.29$, F10214 thereafter), SMM\,J14011+0252 ($z=2.57$, SMM14011
thereafter) and H1413+117 (Cloverleaf Quasar, $z=2.56$). 
We use a $\Lambda$ cosmology with $H_{\rm 0} = 71$
\kms\,Mpc$^{-1}$, $\Omega_\Lambda=0.73$ and $\Omega_m=0.27$ (Spergel
\etal\ \cite{spergel03}).\\

\section{Observations}

\begin{figure*} \centering
\includegraphics[width=18.0cm]{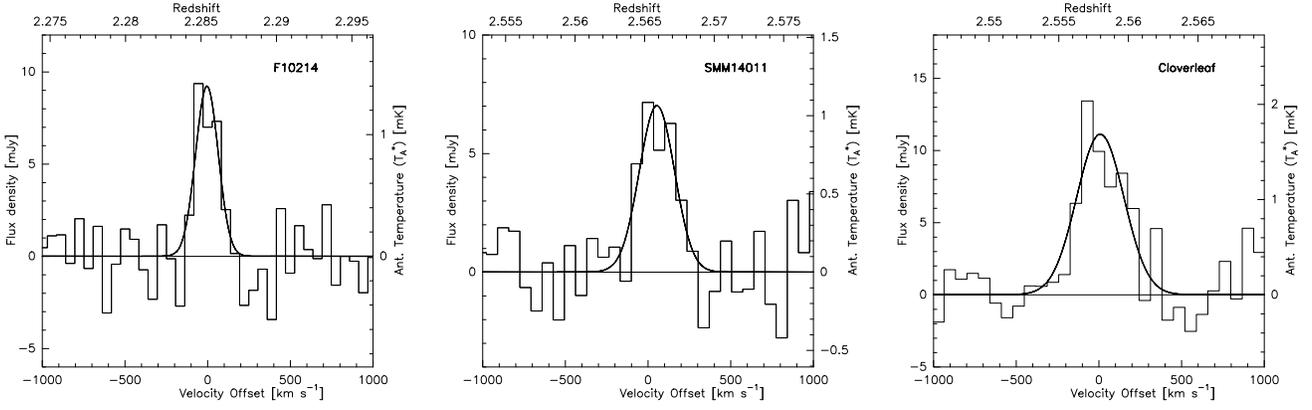}
\caption{Spectra of the \cone\ fine structure line towards
F10214, SMM14011 and the Cloverleaf superposed on their Gaussian fit profiles
(see Table~\ref{linepara} for parameters). The velocity scale is
relative to the CO redshift given in the footnote of
Table~\ref{linepara}. The velocity resolution is 55\,\kms\ (F10214), 65\,\kms (SMM14011) and 70\,\kms\
(Cloverleaf).}  \label{lines} \end{figure*}

Observations were carried out with the IRAM 30\,m telescope from Sep.
2003 to July 2004. We used the CD receiver configuration with the C/D
150 receivers tuned to the \cone\ transition. For each source the
observing frequency of the \cone\ transition is listed in
Table\,\ref{linepara}. The beam size at 140\,GHz is
$\approx\,17''$.  Typical system temperatures were $\approx$\,200\,K
and $\approx$\,270\,K ($T_{\rm A}^*$) during winter and summer respectively. The
observations were carried out in the wobbler switching mode, with a
switching frequency of 0.5\,Hz and a wobbler throw of $50''$ in
azimuth.  Pointing was checked frequently and was found to be stable
within $3''$. Calibration was obtained every 12\,min using the
standard hot/cold--load absorber measurements. The planet Mars was
used for absolute flux calibration. The antenna gain was found to be
consistent with the standard value of 6.6 Jy/K at 140\,GHz. We
estimate the flux density scale to be accurate to about $\pm15$\%.\\
Data were recorded using the 4\,MHz filter banks on each receiver (512
channels, 1GHz bandwidth, 4\,MHz channel spacing). The data were processed
using the CLASS software. After dropping bad data only linear
baselines were subtracted from individual spectra. The resulting
profiles were regridded to a velocity resolution of 
55\,\kms\ (F10214), 65\,\kms (SMM14011) and 70\,\kms\ (Cloverleaf) leading to an RMS of 0.25\,mK
(1.7\,mJy), 0.24\,mK (1.6\,mJy) and 0.25\,mK (1.7\,mJy), respectively.  
The on-source integration time in the final spectra is 8h, 10.5h and 4.5h for 
F10214, SMM14011 and the Cloverleaf respectively. The final spectra
are presented in Fig.~\ref{lines}.

\section{Results} 

\subsection{F10214}

Our new \cone\ spectrum towards F10214 does not confirm the \cone\ detection
reported by Brown \& Vanden Bout (\cite{brown92}, see Fig.\,1 and
Table 1 versus their Fig.\,3). The integrated line flux is 7 times lower and the
peak line antenna temperature of 1.4\,mK is well within the noise
level of the old observations. Given the superior quality of our new
measurements, we consider the data presented here to be the first detection of C{\sc i}
in this object. The \cone\ redshift in F10214 is in agreement
with the CO redshift but the carbon line is narrower. We attribute this to the limited signal
to noise ($\approx$\,6) in our observations. Our derived line parameters are
summarized in Tab.~\ref{linepara}. For F10214 we find a line
luminosity (see e.g. Solomon \etal\ \cite{solomon97} for a definition) ratio
compared to \cco\ of $L'_{\cone}/L'_{\cco} = 0.19 \pm 0.05$ (see
footnote of Table\,\ref{linepara} for \cco\ luminosities and
references).

\subsection{SMM14011}

Our \cone\ detection in SMM14011 is the first atomic carbon detection
reported in any high--$z$ Scuba sub-mm galaxy. The absence of AGN
characteristics in optical spectra (Barger \etal\ \cite{barger99},
Ivison \etal\ \cite{ivison00}) together with the non detection of hard X-rays (Fabian \etal\ \cite{fabian00})
implies that the gas is mainly heated by star formation.
The carbon line linewidth is slightly higher than the CO width and the
spectrum shows a small velocity offset compared to the CO redshift --
this is most likely due to the current limited signal to noise ratio.
The \ci\ to \cco\ line luminosity ratio in SMM14011 is $L'_{\cone}/L'_{\cco} = 0.32 \pm 0.06$.

\subsection {The Cloverleaf}

The Cloverleaf is a Broad-Absorption Line (BAL) quasar (QSO) with very broad
high-excitation emission lines and is the brightest and
best--studied high--$z$ CO source. The redshift of the carbon line agrees with the
CO redshift and the \cone\, linewidth of 360\,\kms\ is similar to CO
measurements (e.g. Wei\ss\ \etal\ \cite{weiss03}) and the previous
neutral carbon detection by Barvainis \etal\ (\cite{barvainis97}). The
integrated \cone\, flux density is in good agreement with
the value reported by Barvainis. With our new \cone\ flux of 
$I_{\rm CI} = 3.9\,\pm\,0.6$\,Jy \kms\ we find a line luminosity ratio
of $L'_{\cone}/L'_{\cco} = 0.15 \pm 0.02$. Our observations confirm the low carbon fine structure
line ratio of $L'_{\ctwo}$/$L'_{\cone} \approx 0.5$ (Wei\ss\ \etal\
\cite{weiss03}).

   \begin{table*}
      \caption[]{Observed \cone\ line parameters. }
         \label{linepara}
            \begin{tabular}{l c c c c c c c c}
            \hline
            \noalign{\smallskip}
            Source & $\nu_{obs}$ &$T_{\rm A}^*$& $S_\nu$ & $\Delta
      V_{\rm FWHM}$ & $I$& $V_0$ & $L'_{\rm CI}$/10$^{10}$ & $^{\mathrm{d}}$$L'_{\rm CI}$/$L'_{\rm CO(3-2)}$ 
            \\
                 & [GHz] &[mK] &[mJy] & [\kms] & [Jy \kms]&[\kms] & [K
      \kms\, pc$^2$] &\\
            \noalign{\smallskip}
            \hline
            \noalign{\smallskip}
            
 F10214           & 149.8024 & 1.4 $\pm$ 0.15 & 9.2 $\pm$ 1.0     & 160 $\pm$ 30
 & 1.6 $ \pm$ 0.2  & $^{\mathrm{a}}$ --5 $\pm$ 12 &2.2 $\pm$ 0.3 &0.19
                     $\pm$ 0.05\\
 SMM14011             & 138.0457 & 1.1 $\pm$ 0.2  & 7.3 $\pm$ 1.5  & 235 $\pm$ 45 
 & 1.8 $\pm$ 0.3  & $^{\mathrm{b}}$ 45 $\pm$ 27 &3.1 $\pm$ 0.5 & 0.32
                     $\pm$ 0.06\\
 Cloverleaf          & 138.3313   & 1.8 $\pm$ 0.3 &  11.2 $ \pm$ 2.0  & 360 $\pm$ 60  
                     & 3.9 $\pm$ 0.6 &  $^{\mathrm{c}}$ 7 $\pm$ 25 
                     & 6.7 $\pm$ 1.0 &0.15 $\pm$ 0.02\\
            \noalign{\smallskip}
            \hline
           \end{tabular}
\begin{list}{}{}
\item[] Quoted errors are statistical errors from Gaussian
            fits. Systematic calibration uncertainty is $\pm\,15$\%  
\item[$^{\mathrm{a}}$] Center velocity relative to z=2.2854 (Downes \etal\ \cite{downes95}).
\item[$^{\mathrm{b}}$] Center velocity relative to z=2.5653 (Frayer \etal\ \cite{frayer99}).
\item[$^{\mathrm{c}}$] Center velocity relative to z=2.5578 (Wei\ss\ \etal\ \cite{weiss03}).
\item[$^{\mathrm{d}}$] $L'_{\rm CO(3-2)}$: F10214:
11.7$\pm$0.6$\times\,10^{10}$ K\,\kms\,pc$^2$ Solomon \etal\  (\cite{solomon92}); SMM14011: 9.7$\pm$1.0$\times\,10^{10}$
K\,\kms\,pc$^2$ Downes \& Solomon (\cite{downes03}); Cloverleaf 45.7$\pm$0.7$\times\,10^{10}$ K\,\kms\,pc$^2$ 
Wei\ss\ \etal\ (\cite{weiss03})

\end{list}
   \end{table*}

\section{Discussion}

\subsection{\ci/CO line ratios and cooling}
\noindent Judging from our $L'_{\ci}/L'_{{\rm CO}(3-2)}$ line luminosity ratios
the carbon chemistry and/or excitation does not change strongly
between a pure starburst (SMM14011) and the starburst/AGN (Cloverleaf,
F10214) heated environment at redshifts $\approx2.5$. For SMM14011 we
find a slightly higher $L'_{\ci}/L'_{{\rm CO}(3-2)}$ line luminosity ratio than for the two QSOs. 
For local galaxies Gerin \& Phillips (\cite{gerin00}) find an average brightness
temperature ratio between \cone\ and \aco\ of $0.2\pm0.2$ independent
of the environment.  A similar ratio has also been found in Arp\,220
(Gerin \& Phillips \cite{gerin98}). Detailed radiative transfer models 
show that the \aco\ and \cco\ transitions are near thermal equilibrium 
($L'_{{\rm CO}(1-0)} \approx L'_{{\rm CO}(3-2)}$) for all 
three sources studied here (Wei\ss\ \etal\ in prep.).  
This implies that the  \cone/\aco\ ratio for dusty quasars and 
submm galaxies at $z\approx\,2.5$ resembles those for local
galaxies. The cooling via \ci\ and CO for all three sources is small compared to
the cooling due to thermal dust emission ($L_{\rm CO}/L_{FIR} \le
10^{-4}$) with CO being the more effective cooler compared to \ci\ ($L_{{\rm
\ci}}/L_{\rm CO} \approx 0.05 - 0.2$).\\

\subsection{Neutral carbon mass and abundance}
\begin{figure} 
\centering
\includegraphics[width=8.5cm]{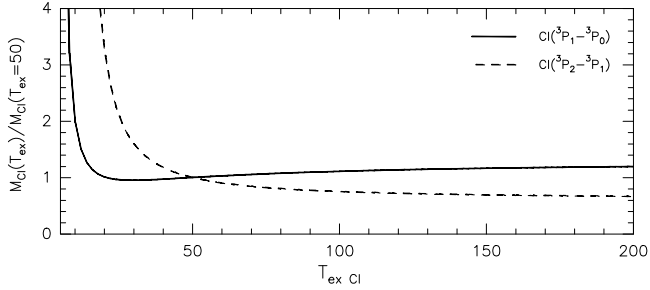} 
\caption{Dependence of the neutral carbon mass on the carbon excitation temperature assuming
LTE. The carbon mass is normalized to $T_{ex\,\ci} = 50$\,K. The solid
line corresponds to masses derived from the \cone\ line luminosity
(Eq.\ \ref{nrmci}), the dashed line is the corresponding plot of the
\ctwo\ line luminosity.}  \label{mci-tex} \end{figure}

In analogy to the formulas given in Wei\ss\ \etal\ (\cite{weiss03})
for the upper fine structure line of atomic carbon, \ctwo, we can  
derive the total mass of neutral carbon using the luminosity of the lower fine structure
 line, \cone, via  
\begin{equation}
\label{eqmci}
M_{\ci} = C\,m_{\ci}\,\frac{8 \pi k \nu_{0}^{2}}{h c^3
A_{10}}\,Q(\tex)\,\frac{1}{3}\,{\rm e}^{T_{1}/
\tex}\,L'_{\cone}
\end{equation}

\noindent where $Q(\tex)=1+3{\rm e}^{-T_{1}/\tex}+5{\rm e}^{-T_{2}/\tex}$ is the
\ci\, partition function. $T_{1}$\,=\,23.6\,K and $T_{2}$\,=\,62.5\,K
are the energies above the ground state. $m_{\ci}$ is the mass of a
single carbon atom and $C$ is the conversion between pc$^2$ and
cm$^2$. The above equation assumes optically thin \ci\ emission and
that both carbon lines are in local thermodynamical equilibrium (LTE).
Inserting numbers yields:
\footnote{In our previous paper (Wei\ss\ \etal\ \cite{weiss03}) the
similar equation for the carbon mass derived from the upper fine
structure line (Eq.~3) should have a coefficient of
$4.556\times10^{-4}$. All the carbon masses quoted in the text,
however, are correct.}
\begin{equation}
\label{nrmci}
M_{\ci} = 5.706\times10^{-4}\,Q(\tex)\,\frac{1}{3}\,e^{23.6/ \tex}\,L'_{\cone} [\msol] 
\end{equation}
The determination of the excitation temperature of atomic carbon requires the measurement of both fine
structure lines. The upper carbon fine structure line so far has only
been detected in the Cloverleaf Quasar (Wei\ss\ \etal\ \cite{weiss03})
yielding $\tex$\,=\,30\,K ($T_{\rm dust}$\,=\,50\,K). As the excitation
temperature is a priori unknown for the other two sources we here
assume \tex=$30$\,K as derived for the Cloverleaf. Similar excitation
temperatures for all three sources are motivated by similar dust temperatures 
(F10214: $T_{\rm dust}$\,=\,55\,K Benford \etal\ \cite{benford99};
SMM14011: $T_{\rm dust}$\,=\,50\,K Ivison
\etal\ \cite{ivison00}). We note however that
the total neutral carbon mass is not a strong function of the assumed
$T_{ex}$ unless the excitation temperature is below 20\,K
(see Fig.~\ref{mci-tex}). The resulting carbon masses 
(uncorrected for the magnification $m$) are listed in Table~\ref{cmass}. 
\noindent Table~\ref{cmass} also lists the \hh\ masses derived from the \cco\
line luminosity. Here we assume a ULIRG conversion factor of $M_{\hh}/L'_{\rm
CO}$\,=\,0.8\,$\msol\, ({\rm K}\kms\,{\rm pc}^2)^{-1}$ (Downes \&
Solomon \cite{downes98}). The mass ratio between \ci\ and \hh\ allows
us to estimate the neutral carbon abundance relative to \hh\ via
$X[\ci]/X[\hh]\,=\,M(\ci)/(6\,M(\hh))$. Note that the carbon abundance
is independent of the magnification (disregarding differential
magnification) and the applied cosmology. We find a carbon abundance
of $X[\ci]/X[\hh]\approx\,5\times10^{-5}$ for all sources. This value
is similar to the \ci\ abundance in the Galaxy of
$X[\ci]/X[\hh]=2.2\times10^{-5}$ (Frerking \etal\ \cite{frerking89}).
This implies that the cold molecular gas in these systems at redshift 2.5 is
already substantially enriched. This finding is in line with abundance 
determinations for high redshift QSOs using optical spectra
(e.g. Dietrich \etal\ \cite{dietrich03}) albeit for the cold molecular gas phase.
For local starburst galaxies it has been found that the
neutral carbon gas phase abundance tends to be higher
than in the Milky Way (e.g. Schilke \etal\ \cite{schilke93}, White
\etal\ \cite{white94}, Israel \& Baas \cite{israel01},
\cite{israel03}). However, given the underlying assumptions of \hh\ mass
determinations in high--$z$ sources, our data do not allow 
us to investigate possible \ci\ abundance variations between QSO and starburst
heated environments in more detail.

\noindent We conclude that the physical properties of systems at large 
lookback times ($z=2.5$ corresponds to an age of the universe of only
2.7\,Gyr) are similar to today's starburst/AGN environments.
Our study also shows that \ci\ is one of the brightest tracers of the
cold molecular gas in galaxies. It therefore provides a powerful
diagnostic of galaxy evolution well beyond $z=2.5$, especially 
in the light of the next generation of mm and sub--mm telescopes (such as ALMA).

\begin{table} 
\caption[]{Mass and abundance of neutral carbon. \label{cmass}} 
\begin{tabular}{l c c c} 
\hline \noalign{\smallskip} &F10214& SMM14011 & Cloverleaf\\
\noalign{\smallskip} \hline \noalign{\smallskip}

 $M_{\ci}$ [\msol/$10^{7}$] & $2.7\,m^{-1}$ & $3.9\,m^{-1}$ & $8.4\,m^{-1}$\\
 $M_{\hh}$ [\msol/$10^{10}$] & $9.4\,m^{-1}$ & $7.8\,m^{-1}$& $36.5\,m^{-1}$\\
 $X[\ci]/X[\hh]$ &4.9\,$\times\,10^{-5}$ &8.3\,$\times\,10^{-5}$ & 3.8\,$\times\,10^{-5}$ \\

          \noalign{\smallskip}
            \hline
           \end{tabular}
\begin{list}{}{}
\item[] $m$ denotes the magnification due to gravitational lensing
\end{list}
\end{table}
\begin{acknowledgements}
 IRAM is supported by
 INSU/CNRS (France), MPG (Germany) and IGN (Spain). 
\end{acknowledgements}

\end{document}